\documentstyle[pra,aps,epsfig,multicol]{revtex}

\begin{document}
\title{Electronic, vibrational and magnetic properties of 
a novel ${\rm C}_{48}{\rm N}_{12}$ aza-fullerene} 
\author{ Rui-Hua Xie and Garnett W. Bryant} 
\address{National Institute of 
Standards and Technology, Gaithersburg, MD 20899-8423, USA}
\author{Vedene H. Smith, Jr.}
\address{Department of Chemistry, Queen's University, Kingston,
ON K7L 3N6, Canada}
\date{\today}
\maketitle
\begin{abstract}

We study the structural, electronic, vibrational and magnetic properties 
of a novel ${\rm C}_{48}{\rm N}_{12}$ aza-fullerene using  density 
functional theory and restricted Hartree-Fock theory. Optimized geometries 
and total energy of this fullerene have been calculated.  We find that 
for ${\rm C}_{48}{\rm N}_{12}$ the total ground state energy is about 
-67617 eV, the HOMO-LUMO gap is about 1.9 eV, five 
strong IR spectral lines are located at the vibrational frequencies, 461.5 ${\rm cm}^{-1}$, 568.4 
${\rm cm}^{-1}$, 579.3 ${\rm cm}^{-1}$, 1236.1 ${\rm cm}^{-1}$, 1338.9 
${\rm cm}^{-1}$,   the Raman scattering activities and 
depolarization ratios are zero, and 10  NMR spectral signals are predicted. 
Calculations of diamagnetic shielding factor, static dipole polarizabilities and 
hyperpolarizabilities of ${\rm C}_{48}{\rm N}_{12}$ are  performed and discussed.  
Our results suggest that ${\rm C}_{48}{\rm N}_{12}$ may 
have potential applications as semiconductor components and 
possible building materials for nanometer electronics, photonic devices  
and diamagnetic superconductors. 

\end{abstract}

\begin{multicols}{2} 

Since the discovery of ${\rm C}_{60}$\cite{kroto}, 
doped fullerenes have attracted a great deal of interest 
in the research community of physics, chemistry and material 
engineering due to their remarkable structural, electronic, 
optical and magnetic properties\cite{book1,book3,book4,book5}. 
For example,  it has been shown that the doped fullerenes can 
exhibit large third-order optical nonlinearities and be ideal candidates 
as photonic devices including all-optical switching, data processing and 
eye and sensor protection\cite{book4}. Another example is 
alkali-doped ${\rm C}_{60}$ crystals, which can become superconductors
\cite{holczer,on99}, for example,  at a critical temperature 
${\rm T}_{\rm c}=30$ K \cite{holczer}. Besides the 
alkali metal doping, there is also another type 
of doping, named {\sl substitute doping}
\cite{book1,book3,book4,book5}, i.e., substituting one 
or more carbon atoms of fullerenes by other atoms. Over the past 
10 years, boron and nitrogen atoms have been successfully used to 
replace carbon atoms of ${\rm C}_{60}$ and synthesize many kinds of
 doped ${\rm C}_{60}$ molecules, 
${\rm C}_{\rm 60-m-n}{\rm X}_{\rm m}{\rm Y}_{\rm n}$
\cite{book1,book3,book4,book5,guo91}. A very efficient 
method of synthesizing ${\rm C}_{59}{\rm N}$ 
has been recently developed by Hummelen {\sl et al.}
\cite{hummelen95}. Their method has led to a number of 
detailed studies of the physical and chemical 
properties of ${\rm C}_{59}{\rm N}$\cite{book3,book4,book5}, which 
suggests  that  the new type of substituted fullerenes may be 
used as semiconductor components and possible building materials 
for future nanometer electronics since their band gaps and electronic 
polarizations can vary with different substitute doping
\cite{book1,book3,book4,book5}. 

Very recently, ${\rm C}_{60}$ with more than one 
nitrogen atom replacing a carbon atom in the cage has been synthesized by  
Hultman {\sl et al.}\cite{hultman01}.  They reported the existence of 
a novel ${\rm C}_{48}{\rm N}_{12}$ aza-fullerene\cite{hultman01,stafstrom},  
which has one nitrogen atom  per pentagon and the symmetry of the 
$S_{6}$ point group. Hence, it would be interesting and significant 
to investigate and predict the structural, electronic, vibrational and 
magnetic properties of this new  aza-fullerene from the viewpoint of 
practical applications. This forms the purpose of the present paper. 
In the following, we report  {\sl ab initio} calculations of the optimized 
geometric structure, electronic properties, vibrational frequencies, 
IR spectra, NMR shielding tensors, diamagnetic shielding factor, static dipole 
polarizabilities and hyperpolarizabilities of the stable 
${\rm C}_{48}{\rm N}_{12}$ aza-fullerene.  Our results suggest 
that ${\rm C}_{48}{\rm N}_{12}$  may have potential applications 
as semiconductor components and possible building materials for 
nanometer electronics and diamagnetic  superconductors.

\centerline{\epsfxsize=4in \epsfbox{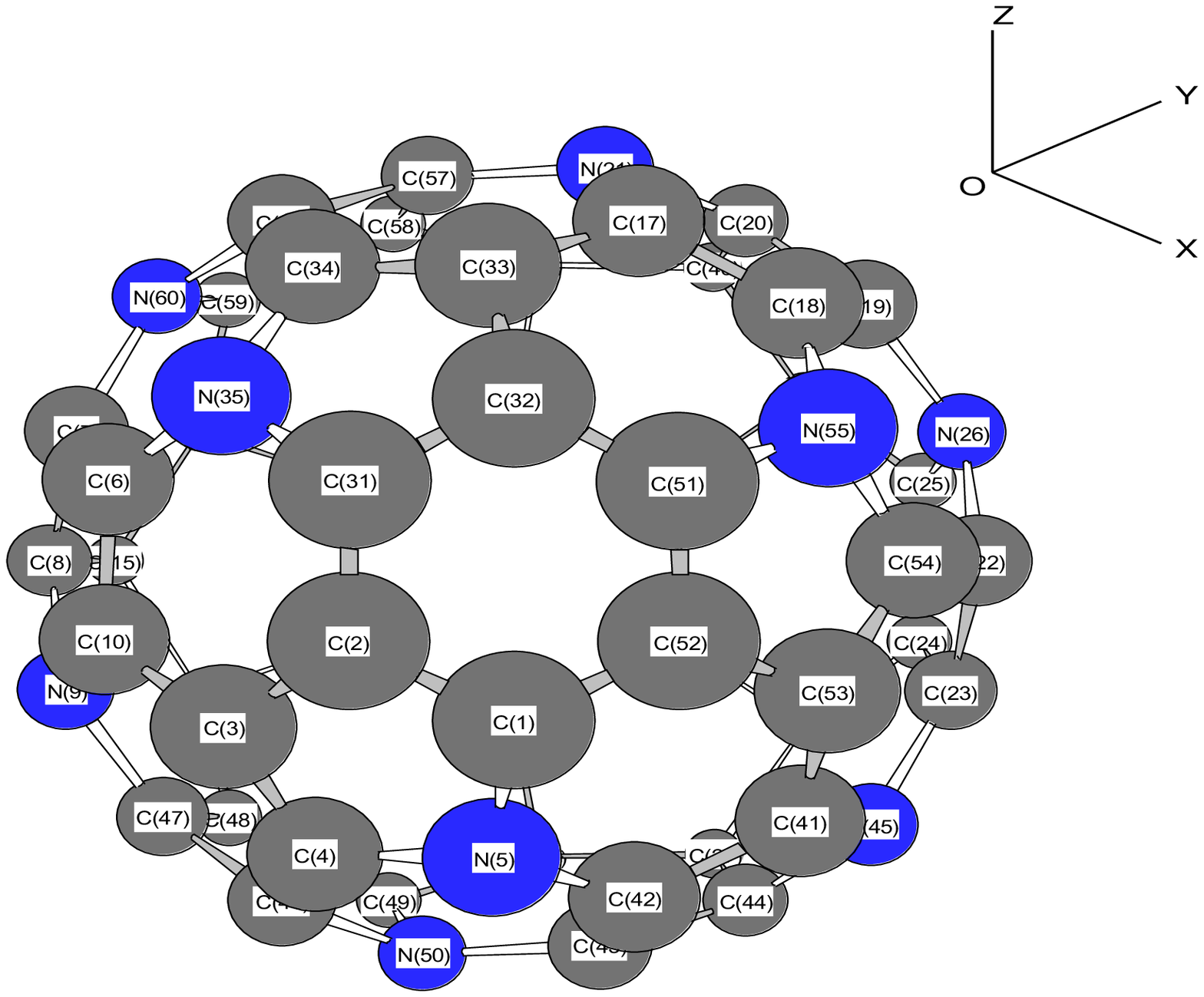}}

\begin{quote}
{\bf FIG.1}: {\small ${\rm C}_{48}{\rm N}_{12}$
 geometry structure obtained from {\sl ab initio}
B3LYP geometry optimizations with a 6-31G basis set. 
The center numbers \{5, 9, 14, 21, 26, 30, 35, 39, 45, 50, 55, 60\} 
are for nitrogen atoms and the others for carbon atoms.}
\end{quote}

Full geometry optimization and total energy calculation of 
${\rm C}_{48}{\rm N}_{12}$ were performed by using the 
NWChem program\cite{nwchem,nist}, where we have employed 
the B3LYP\cite{becke93} hybrid density functional 
theory (DFT) method that includes a mixture of Hartree-Fock 
(exact) exchange, Slater local exchange\cite{slater74}, Becke 88 
non-local exchange\cite{becke88}, the VWN III local 
exchange-correlation functional\cite{vosko80} and
the LYP correlation functional\cite{lyp88}. Throughout the 
split-valence 6-31G basis set is used in our calculations. All 
results are calculated to machine precision. As will be discussed later, 
the best total energy calculations with other basis sets differ 
from the 6-31G basis set results by less than one percent. This 
should define the accuracy of our calculations.

The optimized geometric structure of ${\rm C}_{48}{\rm N}_{12}$ is presented 
in Fig.1. As shown by Hultman et al.\cite{hultman01},  
the {\sl ab initio} calculations show that there is only one nitrogen 
atom per pentagon  and two nitrogen atoms preferentially sit 
together in one hexagon. The symmetry of the stable isomer of   
${\rm C}_{48}{\rm N}_{12}$ is the $S_{6}$ point group\cite{stafstrom}.
The radial distance ${\rm R}_{i}$ from the $i$th atom to  the center 
(i.e., the origin) of ${\rm C}_{48}{\rm N}_{12}$ is listed in Table I. 
We find that there are 10 different radii, which suggest that 
${\rm C}_{48}{\rm N}_{12}$ is an ellipsoid structure, different from 
${\rm C}_{60}$ where each carbon atom has equal radial distance from 
the center.

\end{multicols}
 
\begin{quote}
{\bf Table I:}  Calculated net Mulliken charges ${\rm Q}_{\rm i}$ ( $ 
q = 1.602\times 10^{-19}$), electrostatic 
potential ${\rm V}_{\rm i}$ ( $v =2.7209\times 10^{3}$), 
electron density $\rho_{i}$ ($\eta =  1.08\times 10^{12})$) 
 and radial distance ${\rm R}_{\rm i}$ at the atom center number ${\rm n}_{i}$  
in ${\rm C}_{48}{\rm N}_{12}$ by using B3LYP  hybrid DFT method with a 
split-valence 6-31G basis set.    
\end{quote}
\begin{center}
\begin{tabular}{lccccc}\hline\hline
$\ \ \ \ \ \ \ \ \ {\rm n}_{i}$  & Atom & ${\rm R}_{\rm i}$   
&  ${\rm Q}_{\rm i}/q$  & ${\rm V}_{\rm i}/v$ &$\rho_{i}/\eta$   \\ 
            &                & [ ${\rm nm}$] & [ ${\rm C}$ ] & [ ${\rm V}$ ]  & [ ${\rm C/m}^{3}$ ] \\ \hline
\{ 8, 20, 22, 44, 46, 56\}  & C &0.3495 &  0.262 & -14.6663 & 117.999\\
\{ 10, 17, 25, 41, 49, 59\} & C &0.3508 &  0.255 & -14.6674 & 118.029\\
\{ 4, 15, 27, 34, 40, 54\}  & C &0.3541&  0.249 & -14.6679 & 118.027\\
\{1, 13, 16, 31, 38, 51\}   & C &0.3535&  0.215 & -14.6686 & 118.058\\
\{ 7, 19, 23, 43, 47, 57\}  & C &0.3448&  0.210 & -14.6631 & 118.004\\
\{ 6, 18, 24, 42, 48, 58\}  & C &0.3470&  0.199 & -14.6694 & 117.992\\
\{ 2, 12, 29, 32, 37, 52\}  & C &0.3545&  0.002 & -14.7041 & 118.109\\
\{  3, 11, 28, 33, 36, 53\} & C &0.3549& -0.017 & -14.7026 & 118.103\\
\{ 5, 14, 30, 35, 39, 55\}  & N &0.3595& -0.669 & -18.2639 & 191.331\\
\{9, 21, 26, 45, 50, 60\}   & N &0.3531& -0.705 & -18.2547 & 191.325\\
Carbon atoms in ${\rm C}_{60}$ & C & 0.3535 & 0 & -14.7146 & 118.181 \\ \hline
\end{tabular}
\end{center}

\begin{multicols}{2}

The net Mulliken charges of carbon and nitrogen atoms in 
${\rm C}_{48}{\rm N}_{12}$ are listed in Table I. It is found that 
there are two types of nitrogen atoms in the structure. The net 
Mulliken charges on the two types of nitrogen atoms are $-0.67\ {\rm q}$ and 
$-0.71\ {\rm q}$. The carbon atoms separate into two groups, 
one-fourth of the carbon atoms having net Mulliken charges in the range 
of $-0.02\ {\rm q}$ to $0.01\ {\rm q}$ and the remaining three-fourth in 
the range of $0.19\ {\rm q}$ to $0.27\ {\rm q}$. Although the Mulliken 
analysis cannot estimate the atomic charges quantitatively, their 
signs can be estimated\cite{szabo82}.  From these results, we find that 
the doped nitrogen atom exists as an electron acceptor whose net Mulliken 
charge is negative, and 7/8  of the carbon atoms as  electron donors whose 
net Mulliken  charges are positive. We should mention that we also 
performed calculations of net Mulliken charges of carbon and boron atoms 
 in ${\rm C}_{48}{\rm B}_{12}$. It is found that the doped boron atom exists
 as an electron donor and  carbon atom as an electron acceptor (more detailed 
results for ${\rm C}_{48}{\rm B}_{12}$ will be reported in another paper\cite{rhxie03}), 
which is consistent with the experimental result\cite{guo91}. Therefore,  
${\rm C}_{48}{\rm N}_{12}$ and ${\rm C}_{48}{\rm B}_{12}$ have opposite 
electronic polarizations, while ${\rm C}_{60}$ is isotropic. In the case of 
doping into silicon, we know that the doped phosphorous 
(the V family in the periodic table) exists as a donor, 
while the doped boron (the III family) exists as an 
electron acceptor. Thus, the results for ${\rm C}_{48}{\rm N}_{12}$ and 
${\rm C}_{48}{\rm B}_{12}$ differ greatly from that for silicon.  We propose 
that ${\rm C}_{48}{\rm N}_{12}$ and ${\rm C}_{48}{\rm B}_{12}$ may be important 
components of semiconductors with opposite electronic polarizations.  

The optimized C-C and  C-N bond lengths in ${\rm C}_{48}{\rm N}_{12}$ are
listed in Table II. We find that there are six different C-N bond
lengths and nine different C-C bond lengths, which are in agreement with
the calculations of Stafstr\"{o}m {\sl et al.}\cite{stafstrom}.
The calculated C-C bond lengths in ${\rm C}_{60}$ are also shown in
Table II. It is seen that ${\rm C}_{60}$ has only two kinds of bond lengths,
i.e., one single C-C bond with $0.14445\ {\rm nm}$  and one double
bond with $0.13944\ {\rm nm}$, which are in good agreement with
the results measured by neutron scattering ($0.145\  {\rm nm}$ and
$0.1391\  {\rm nm}$) \cite{david91} or
nuclear magnetic resonance ( $0.146\ {\rm nm}$ and $0.140\ {\rm nm}$ )
\cite{johnson90}. It is obvious that
most of C-C bond lengths in ${\rm C}_{48}{\rm N}_{12}$, due to the
electronic polarizations, are different from those in ${\rm C}_{60}$.
 
The distribution of bond angles (C-C-C, C-N-C,
C-C-N) in ${\rm C}_{48}{\rm N}_{12}$
is shown in Fig.2. The C-C-C  and C-N-C (and C-C-N) bond angles
separate into three groups: one in the range of $107^{\rm o}$
to $109^{\rm o}$, one in the range of $116^{\rm o}$ to $119^{\rm o}$
and the other one in the range of $119.5^{\rm o}$ to $122^{\rm o}$.
In contrast, there are only two kinds of bond angles, $108^{\rm o}$
and $120^{\rm o}$,  in ${\rm C}_{60}$. Hence, only few bond angles
in ${\rm C}_{48}{\rm N}_{12}$ are similar to those of ${\rm C}_{60}$.

\begin{quote}
{\bf Table II:}  Calculated bond lengths in ${\rm C}_{48}{\rm N}_{12}$ 
by using B3LYP  hybrid DFT method with a split-valence 6-31G basis set. 
\end{quote}
\begin{center}
\begin{tabular}{clc}\hline\hline
Bond &  \ \ \ Center Number        &   Bond Length  \\ 
          & \ \ \ \ \ $(n_{i},n_{j})$     & [ ${\rm nm}$ ]  \\  \hline
C-C & (3,     4)  
(11,    15) 
(27,    28) & 0.13971 \\ 
 & (33,    34) 
(36,    40) 
(53,    54) &  \\
C-C & (6 ,   10)  
(17,    18) 
(24,    25)  & 0.14004 \\  
 &(58,    59)  
(41,    42)  
(48,    49)  &  \\
C-C  &(7 ,    8 )  
(19,    20)  
(22,    23)  & 0.14068 \\  
& (43,    44)  
(46,    47)  
(56,    57)  &  \\
C-C &(1 ,     2)   
(12,    13)    
(16,    29)  & 0.14125 \\       
& (31,    32)    
(37,    38)    
(51,    52)  & \\
C-C  &(1 ,    52)   
(2 ,   31)   
(12,    38) & 0.14216 \\ 
& (13,    29)  
(16,    37)  
(32,    51)  &  \\
C-C  &(6 ,    7)   
(18,    19)  
(23,    24) & 0.14217\\  
& (42,    43)  
(47,    48)  
(57,    58)  & \\
C-C  &( 3,    10)   
(11,    59)   
(17,    33)  & 0.14346 \\  
& (25,    36)   
(28,    49)   
(41,    53)   &  \\
C-C & ( 4,    46)  
( 8,    15)  
(20,    40) & 0.14354\\ 
& (22,    54)   
(27,    44)   
(34,    56)   & \\
C-C & ( 2,     3)   
(11,    12)   
(28,    29)   & 0.14488\\ 
&(32,    33)   
(36,    37)   
(52,    53)   & \\
N-C  &(7,    60)
(9,    47)
(19,    26) & 0.14189 \\
& (21,    57)
(23,    45)
(43,    50)  &  \\
N-C & (8 ,    9)   
(20,    21)  
(22,    26) & 0.14149\\  
&(44,    45)  
(46,    50)  
(56,    60)   & \\
N-C & (9 ,   10)   
(17,    21)  
(25,    26)   & 0.14226\\ 
& (41,    45)   
(49,    50)   
(59,    60)  & \\
N-C & (4 ,    5)   
(14,    15)  
(27,    30)  & 0.14286 \\  
&(39,    40)  
(34,    35)  
(54,    55)  & \\
N-C & (1 ,     5)   
(13,    14)   
(16,    30)  & 0.14315\\    
& (31,    35)    
(38,    39)    
(51,    55)   & \\ 
N-C & ( 5,    42)    
( 6,    35)     
(14,    48) & 0.14317\\       
&(18,    55)   
(24,    30)     
(39,    58)   & \\
C-C & carbon in ${\rm C}_{60}$ & 0.13944 \\
C-C & carbon in ${\rm C}_{60}$ & 0.14445 \\  \hline 
\end{tabular}
\end{center}

\begin{center}
\epsfig{file=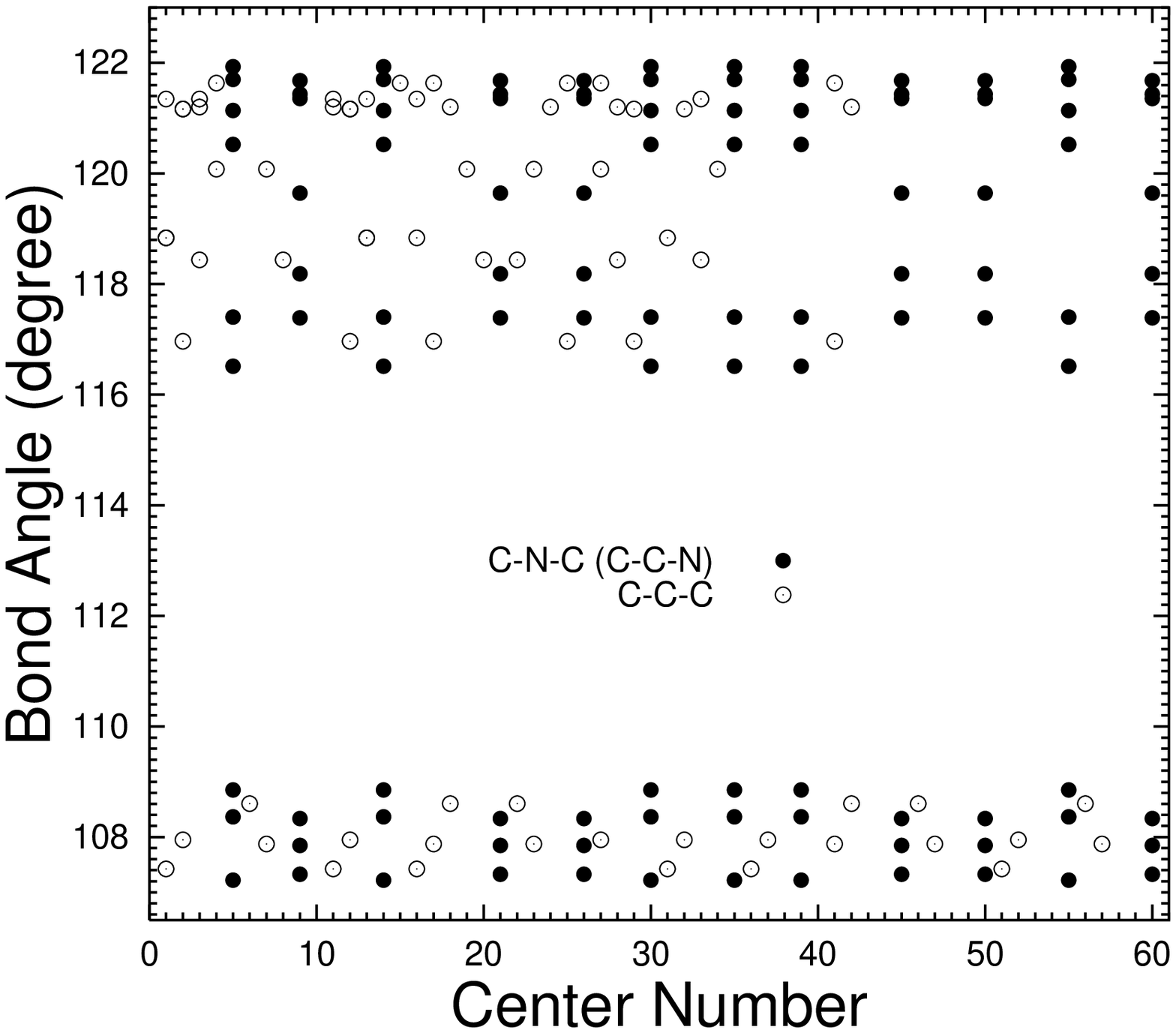,width=6cm,height=6cm}
\end{center}
 
\begin{quote}
{\bf FIG.2}:  {\sl ab initio} B3LYP/6-31G  DFT calculation 
of C-C-C (open circles) and C-N-C (or C-C-N)  (filled circles) 
bond angles in ${\rm C}_{48 }{\rm N}_{12}$. The center numbers 
are labeled in Fig.1.
\end{quote}

\begin{quote}
{\bf Table III:} Total  energy calculations 
 of ${\rm C}_{48}{\rm N}_{12}$
aza-fullerene and ${\rm C}_{60}$ by using B3LYP  hybrid DFT method with a
split-valence 6-31G basis set.  
\end{quote}
\begin{center}
\begin{tabular}{lrr}\hline\hline
        & ${\rm C}_{60}$ \ \ \   & ${\rm C}_{48}{\rm N}_{12}$ \ \ \  \\ \hline
total DFT energy (eV)       & -62192.890        & -67617.316 \\
one electron energy (eV)    & -537319.738       & -577022.480  \\
Coulomb energy (eV)         & 256457.119       & 274997.383 \\
exchange-correlation energy (eV) & -8887.049   & -9361.705 \\
nuclear repulsion energy (eV)    & 227556.778   & 243769.848 \\
LUMO  (eV)                       & -3.382       & -2.707 \\
HOMO  (eV)                       & -6.216       & -4.633 \\
binding energy (eV/atom)         & 8.676        & 8.188  \\
first ionization energy (eV)    &  7.548       &  5.993     \\ \hline
\end{tabular}
\end{center}

The results of total energy calculations of ${\rm C}_{48}{\rm N}_{12}$
aza-fullerene are summarized  in Table III. For comparison, the results
of  ${\rm C}_{60}$ are also listed in Table III. Their DFT  orbital energies with 
 the orbital symmetries are shown in Fig.3.
Because of the valency of the doped nitrogen atoms, the electronic
properties of ${\rm C}_{48}{\rm N}_{12}$ are greatly different from
${\rm C}_{60}$. Since the icosahedral symmetry
of ${\rm C}_{60}$ is lost by the substituted doping, each energy level in
${\rm C}_{48}{\rm N}_{12}$ splits. In detail, we find that the ground state
energy of ${\rm C}_{48}{\rm N}_{12}$ is -67617.316 eV,  
which is 5423 eV lower than that of ${\rm C}_{60}$. For 
${\rm C}_{60}$, the  gap obtained from the onset of the occupied
fivefold-degenerate $h_{u}$ energy level (i.e., HOMO) and the threefold-degenerate
$t_{1u}$ level (i.e., LUMO) is 2.834 eV,
which is in agreement with the
experiment\cite{lof}. For ${\rm C}_{48}{\rm N}_{12}$, the HOMO is a
doubly degenerate level of $a_{g}$ symmetry and the LUMO is a single
level of $a_{u}$ symmetry, and  the corresponding LUMO-HOMO gap is 1.926 eV,  
which is 0.908 eV smaller than
that of ${\rm C}_{60}$. These results indicate the possibility that
${\rm C}_{48}{\rm N}_{12}$ can be the components of
semiconductors with small band gaps. Since ${\rm C}_{48}{\rm N}_{12}$
is isoelectronic with ${\rm C}_{60}^{-12}$, it
corresponds to a complete filling of the $t_{1u}$ and
$t_{1g}$ levels of ${\rm C}_{60}$\cite{hultman01}.  From Fig.3, it is seen that
the HOMO of ${\rm C}_{48}{\rm N}_{12}$ appears at about 1.583 eV above that of
${\rm C}_{60}$.  This difference in the HOMO levels is in agreement with the
calculation of Stafstr\"{o}m  {\sl et al.} \cite{stafstrom} and
corresponds to the difference between the first ionization energies of
${\rm C}_{48}{\rm N}_{12}$ and ${\rm C}_{60}$. These
results  suggested that the aza-fulerene
${\rm C}_{48}{\rm N}_{12}$ is a very good electron donor\cite{stafstrom}.
The binding energies (i.e.,  the difference in the total energies of the
whole system and the fragments) are also listed in Table III.
We find that the binding energy for ${\rm C}_{48}{\rm N}_{12}$ is
8.188 eV per atom, which is about 0.5 eV smaller
than that (=8.676 eV per atom) of ${\rm C}_{60}$. Hence, as synthesized by
Hultman {\sl et al.}\cite{hultman01}, ${\rm C}_{48}{\rm N}_{12}$ would be a
stable aza-fullerene.

\begin{center}
\epsfig{file=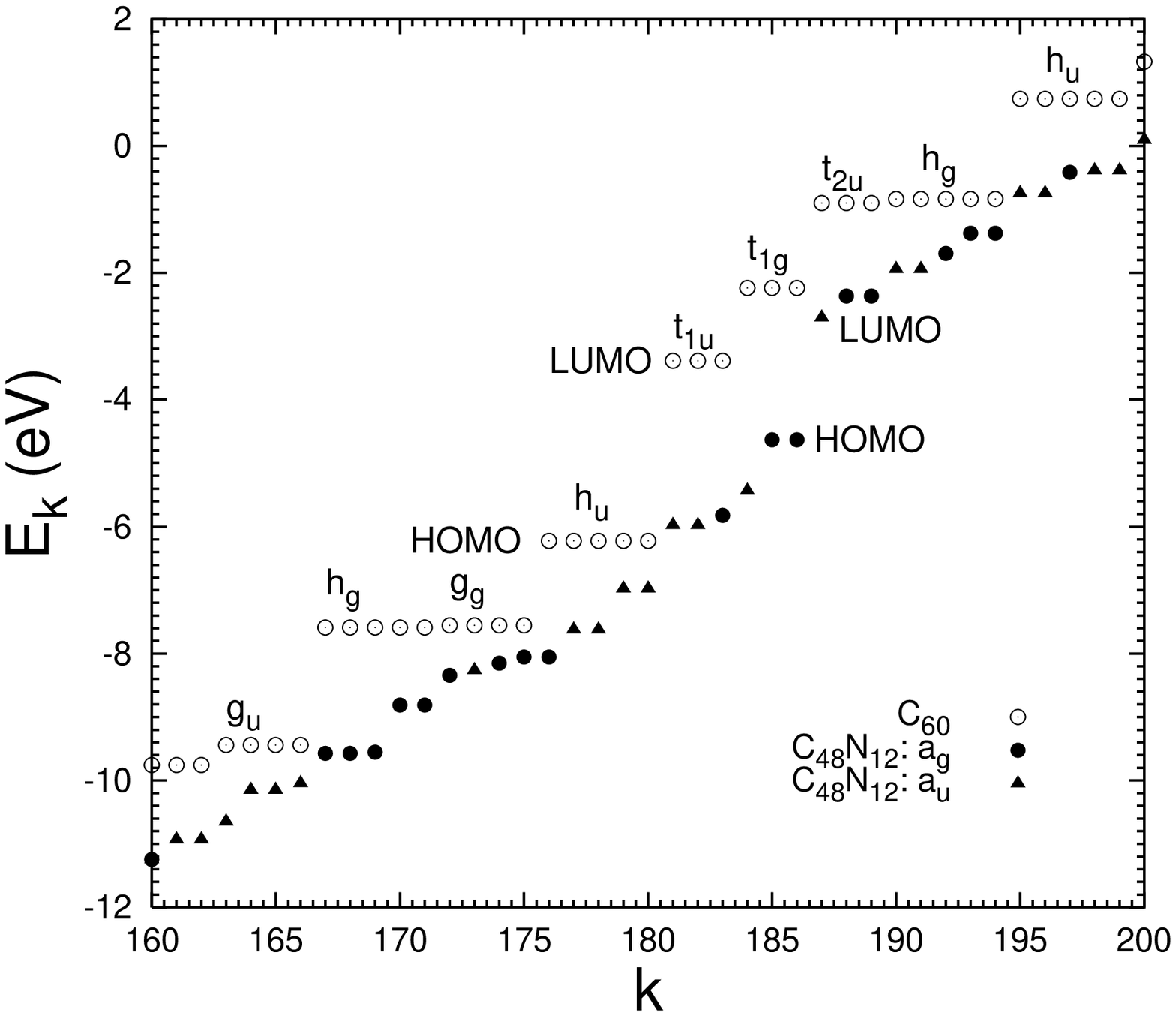,width=7cm,height=6cm}
\end{center}
 
\begin{quote}
{\bf Fig.3}: B3LYP/6-31G DFT orbital energies ${\rm E}_{k}$ at the
$k$th eigenstate. Open circles with labeled  orbital 
symmetries are for ${\rm C}_{60}$. Filled circles with 
orbital symmetry $a_{g}$ and filled triangles with orbital symmetry 
$a_{u}$ are for ${\rm C}_{48 }{\rm N}_{12}$. LUMO and HOMO are shown.  
\end{quote}

The interaction energy of a nucleus of magnetic moment $\mu$ with a
weak external magnetic field ${\bf B}$ is 
given by ${\rm E} = -\mu\bullet {\bf B} (1-\sigma)$, where $\sigma$ 
is the diamagnetic shielding factor (DSF)
\cite{lamb41,feiock} arising from induced magnetic fields at the nucleus.
In Fig.4, we present the calculated  DSF $\sigma$ of carbon and nitrogen
atoms in ${\rm C}_{48}{\rm N}_{12}$ by using 
the NWchem program\cite{nwchem,nist}
with the B3LYP hybrid DFT method and the split-valence 6-31G basis set. In
comparison with the calculated results of ${\rm C}_{60}$ shown in Fig.4, we
find that the average DSF $\overline{\sigma_{c}}$ of carbon atoms
in ${\rm C}_{48}{\rm N}_{12}$ is enhanced by about 3 \%. This enhancement
can be explained as follows. From Table I, we find that the electron density
at each carbon atom in  ${\rm C}_{48}{\rm N}_{12}$, in comparison with
the ${\rm C}_{60}$'s  which is evenly distributed, is decreased due to the
substitute doping.  Correspondingly, as shown in Table I, 
the electrostatic potential at each carbon atom in  ${\rm C}_{48}{\rm N}_{12}$
is also increased. The Lamb formula\cite{lamb41} shows that the 
DSF $\sigma$ is proportional to the electrostatic potential $V$ produced at
the nucleus by the electrons. Hence, the DSF  $\sigma$ of carbon atoms in
${\rm C}_{48}{\rm N}_{12}$ is enhanced. Our calculated 
results  suggest that if ${\rm C}_{48}{\rm N}_{12}$ is doped 
with alkali metal \cite{holczer,on99} and used as a building 
compound for a superconductor material, then it may be
a good diamagnetic superconductor because of the enhanced  DSF in the 
carbon atom.

\begin{center}
\epsfig{file=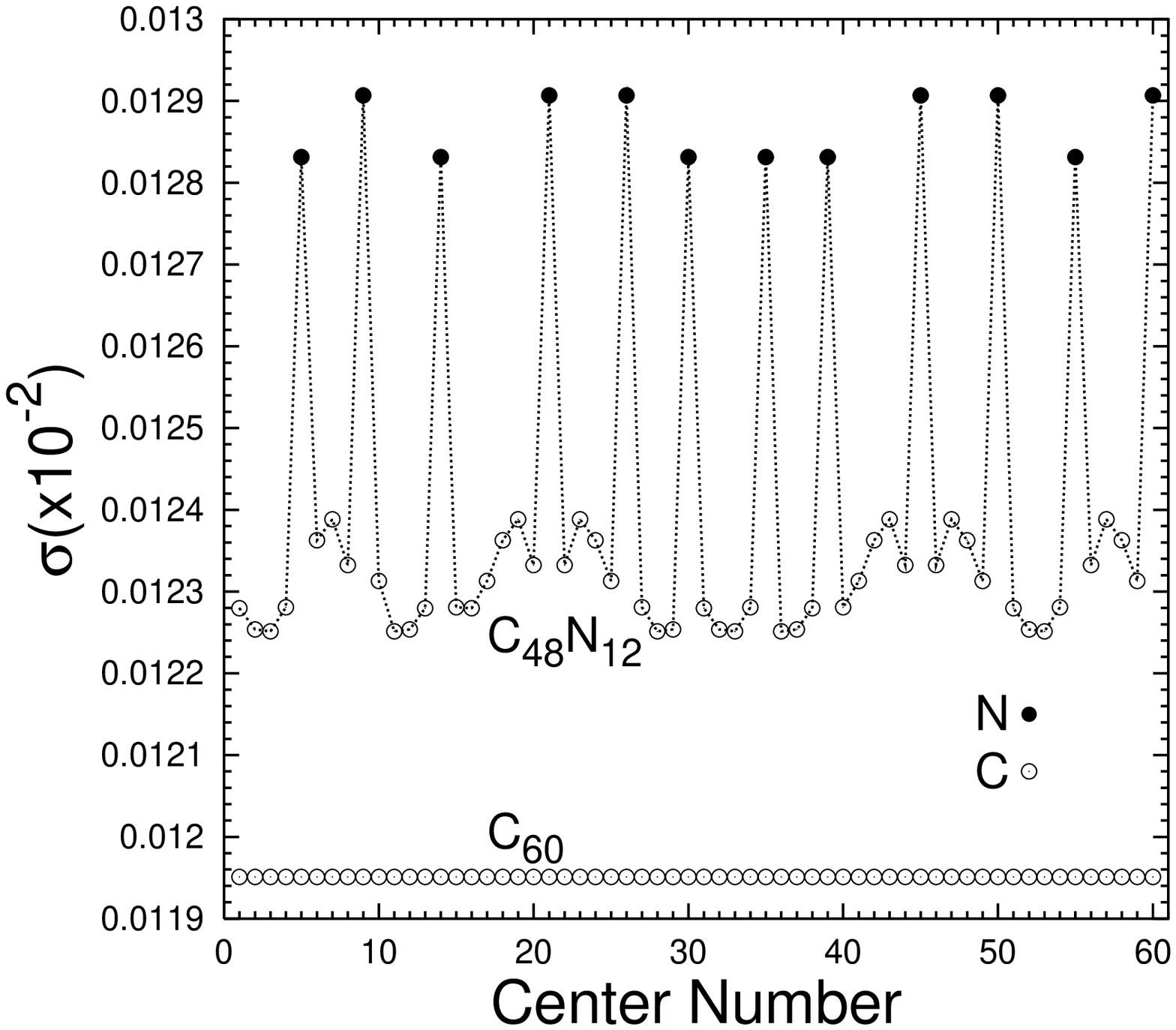,width=7cm,height=6cm}
\end{center}
\begin{quote}
{\bf Fig.4}:  Calculated diamagnetic shielding factor $\sigma$ of
carbon (open circles) and nitrogen (filled circles) atoms in both 
${\rm C}_{60}$ and ${\rm C}_{48 }{\rm N}_{12}$. The center numbers are
labeled in Fig.1.
\end{quote}

Next, we performed  calculations of vibrational frequencies, IR spectrum
, Raman scattering activities and depolarization ratios, and NMR 
shielding tensors  of ${\rm C}_{48}{\rm N}_{12}$ by using the
Gaussian 98 program\cite{nist,gaussian} with B3LYP/6-31G hybrid DFT
method. For ${\rm C}_{48}{\rm N}_{12}$, there are  3n-6 = 174 
vibrational modes for  n=60 (the number of atoms). The 
corresponding vibrational frequencies at  298.15 K and 1 
atmosphere of pressure  are shown in Fig.5, which
also  presents  the IR intensity at each vibrational frequency. 
Since experimental IR spectroscopic data do not directly
indicate the specific type of nuclear motion producing each IR peak, we
do not give here the normal mode information for each vibrational
frequency and the displacements of the nuclei corresponding to the normal
mode. From Fig.5,  we find that ${\rm C}_{48}{\rm N}_{12}$ has
five strong IR lines with the intensities $8.44\times 10^{4}\ {\rm m/mol}$, 
$6.81\times 10^{4}\ {\rm m/mol}$, $5.93\times 10^{4}\ {\rm m/mol}$, 
$5.15\times 10^{4}\ {\rm m/mol}$ and $4.2\times 10^{4}\ {\rm m/mol}$, 
being located at the vibrational frequencies 461.5 ${\rm cm}^{-1}$, 
1236.1 ${\rm cm}^{-1}$, 1338.9 ${\rm cm}^{-1}$, 579.3 ${\rm cm}^{-1}$ 
and 568.4 ${\rm cm}^{-1}$, respectively. The vibrational frequencies, 
461.5 ${\rm cm}^{-1}$, 1236.1 ${\rm cm}^{-1}$, 
1338.9 ${\rm cm}^{-1}$, and 568.4 ${\rm cm}^{-1}$, correspond to two 
degenerate normal modes. Also, it is found that the zero-point correction 
to the total electronic energy is about 10.1 eV, and the thermal 
corrections to the total electronic energy, enthalpy, Gibbs free energy are 10.7 eV, 
10.8 eV and 9.0 eV, respectively.  
 By contrast, the calculated result for
${\rm C}_{60}$ is shown in the inset in Fig.5, where we observe four strong
IR spectral lines with the intensities $2.76\times 10^{4}\ {\rm m/mol}$,
$8.2\times 10^{3}\ {\rm m/mol}$, $1.0\times 10^{4}\ {\rm m/mol}$ and
$1.72\times 10^{4}\ {\rm m/mol}$ at the vibrational frequencies
553.0 ${\rm cm}^{-1}$, 586.5 ${\rm cm}^{-1}$, 1208.5 ${\rm cm}^{-1}$ and
1479.3 ${\rm cm}^{-1}$, respectively, which are in good agreement with
the experiment (528 ${\rm cm}^{-1}$, 577 ${\rm cm}^{-1}$,
1183 ${\rm cm}^{-1}$ and 1429 ${\rm cm}^{-1}$) measured by
Kr\"{a}tschmer {\sl et al.}\cite{wk90}.

\begin{center}
\epsfig{file=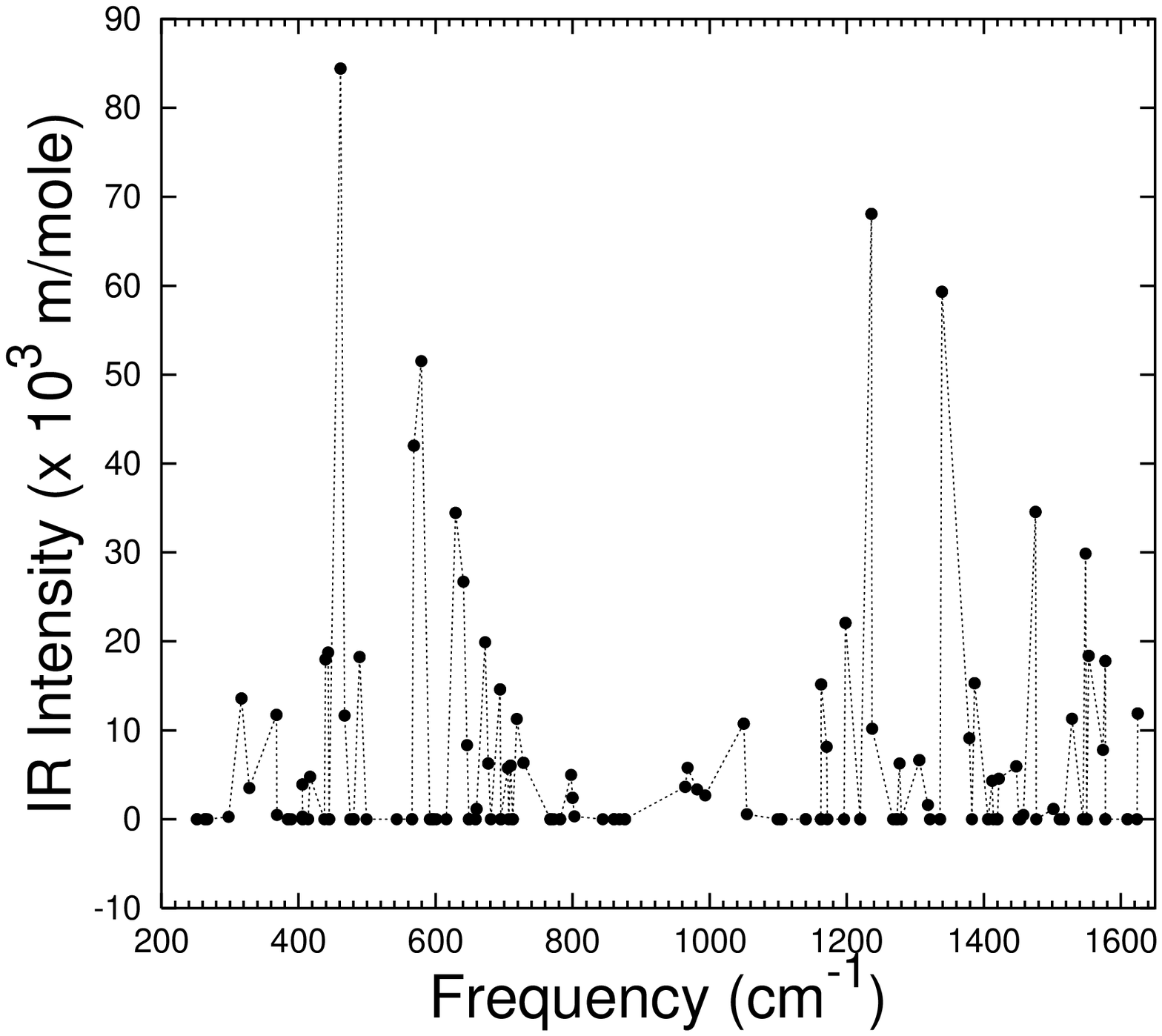,width=6cm,height=6cm}
\end{center}
 
\begin{quote}
{\bf FIG.5}:  {\sl ab initio} B3LYP DFT calculation of
vibrational frequency and IR intensity of
${\rm C}_{48 }{\rm N}_{12}$ with a split-valence
6-31G basis set. Inset is the result for 
${\rm C}_{60}$ calculated with B3LYP/6-31G.
\end{quote}
 
\begin{quote}
{\bf Table IV:}  Static dipole polarizabilities $\alpha$  
( $1\ {\rm au} =0.5292\times 10^{-30}\ {\rm m}^{3}$)  of
${\rm C}_{48}{\rm N}_{12}$ and ${\rm C}_{60}$ by using B3LYP 
hybrid DFT method with a split-valence  6-31G basis set. 
\end{quote}
\begin{center}
\begin{tabular}{ccccccc}\hline\hline
Fullerene        & $\alpha_{xx}$  & $\alpha_{xy}$  & $\alpha_{yy}$ &
 $\alpha_{xz}$  & $\alpha_{yz}$  & $\alpha_{zz}$    \\ 
 & [ au ] & [ au ] & [ au ] & [ au ] & [ au ] & [ au ]  \\ \hline
${\rm C}_{60}$             & 460.84 & 0.00 & 460.84 & 0.00 &0.00 & 460.84  \\
${\rm C}_{48}{\rm N}_{12}$ & 442.55 & 0.00 & 442.54 & 0.00 &0.00 & 445.88 \\ \hline
\end{tabular}
\end{center}

The static dipole polarizability measures the ability of the
valence electrons to find an equilibrium configuration which screens a
static external field\cite{bianchetti}. Hence, molecules with many
delocalized valence electrons should display large values of static
dipole polarizabilities.  In Table IV, we list the calculated
static dipole polarizabilities of  ${\rm C}_{48}{\rm N}_{12}$. For comparison,
the results of ${\rm C}_{60}$ are also listed in Table IV.  Obviously,
large values of static dipole polarizabilities are found for both
${\rm C}_{48}{\rm N}_{12}$ and ${\rm C}_{60}$. Actually, this is expected
since the {\sl sp} structure for  carbon chains 
is rich of $\pi$ bonds delocalized along the entire body 
of the system\cite{bianchetti}. Our calculated
polarizabilities for ${\rm C}_{60}$ are in agreement with those of Fowler
{\sl et al.}\cite{fowler}. From Table IV, we notice that only $\alpha_{xx}$, 
$\alpha_{yy}$ and  $\alpha_{zz}$ are different from zero, i.e., there exist 
three mutually perpendicular directions in the molecule for which the 
induced dipole moments are parallel to the electric field. The locus of points 
formed by plotting $1/\sqrt{\alpha}$ in any direction from the origin yields 
a surface called the polarizability ellipsoid\cite{book8} 
whose axes are x, y, and z. 
Obviously, because the electron distribution for  ${\rm C}_{60}$, as 
discussed before, is isotropic in the x, y and z directions, there is no 
difference between z and x (or y) components of the polarizabilities $\alpha$  
and the polarizability ellipsoid becomes a sphere. 
However, this difference for ${\rm C}_{48}{\rm N}_{12}$ 
is pronounced because of the electron polarization caused by 
the electron acceptor, N atom. The polarizability of 
${\rm C}_{48}{\rm N}_{12}$  
is the same in both x and y directions. Thus, the polarizability ellipsoid in 
${\rm C}_{48}{\rm N}_{12}$ becomes a rotational ellipsoid with two equal 
axes x and y. 

It is known that the molecular vibration must be accompanied 
by a change in the polarizability of 
the molecule in order for a molecular vibration to be Raman active\cite{book8}. 
Although the polarizability ellipsoid may have higher symmetry than 
the molecule,  all the symmetry elements possessed by the molecule will also be 
possessed by the ellipsoid\cite{book8}. Hence,  if the polarizability ellipsoid 
is changed in size, shape, or orientation as a result of vibrational motion, a 
Raman spectrum will result\cite{book8}. In our case,  the calculated Raman 
scattering activities and depolarization ratios for  ${\rm C}_{48}{\rm N}_{12}$ 
are zero. Thus, its vibration is not Raman active, which implies that
the polarizability of the molecule does not change in a molecular vibration, 
and a measurement of its depolarization ratio cannot provide a means of 
distinguishing totally symmetrical vibrations from the rest.

The first-order hyperpolarizability $\beta$ \cite{book9} 
of ${\rm C}_{48}{\rm N}_{12}$ is 
also calculated and found to be zero, the same as  that 
of $C_{60}$. This is actually 
expected since they display inversion symmetries.  Consequently, this 
aza-fullerene ${\rm C}_{48}{\rm N}_{12}$ cannot produce the second-order 
nonlinear optical (NLO) interactions. Based on the available studies of the 
third-order optical nonlinearities of ${\rm C}_{59}{\rm N}$ and 
${\rm C}_{58}{\rm N}_{2}$\cite{book4}, we expect that 
${\rm C}_{48}{\rm N}_{12}$ may exhibit large second-order  
hyperpolarizability $\gamma$ because of its large electronic polarizations 
and small HOMO-LUMO gap.  This suggests that ${\rm C}_{48}{\rm N}_{12}$ may be 
one of the molecules, for which we are looking to build third-order NLO 
materials for photonic applications. 

Modern high-field, multipulse NMR spectroscopy has proven to be an
exceptionally powerful technique in characterizing molecular systems 
and structures. Here we calculate the NMR shielding tensor of carbon and 
nitrogen atoms in ${\rm C}_{48}{\rm N}_{12}$ aza-fullerene. 
 Since  no current functionals include a magnetic field dependence, the DFT
methods do not provide systematically better NMR CS results than
Hartree-Fock theory\cite{gaussian}. Hence, we compute the NMR shielding tensor 
of ${\rm C}_{48}{\rm N}_{12}$ with the gauge-including atomic 
orbital (GIAO) method\cite{giao} by using Gaussian 98 program 
\cite{gaussian} in the level of restricted Hartree-Fock (RHF) theory 
with the split-valence  6-31G basis set. Table V summarizes the results
for the isotropic and anisotropic absolute shielding constants of the carbon 
and nitrogen atoms in ${\rm C}_{48}{\rm N}_{12}$. 
The  isotropic shielding constant $\Delta_{iso}$ is defined as 
$\Delta_{iso} = (\Delta_{xx}+\Delta_{yy}+\Delta_{zz})/3$\cite{jrc96}, 
where $\Delta_{ij}$ (i,j = x, y, z) is the component of the  shielding tensor.  
The anisotropic shielding constant $\Delta_{aniso}$, an indication of the 
quality of the shielding tensor,  is defined  as 
$\Delta_{aniso} = \Delta_{3} - ( \Delta_{1} + \Delta_{2} )/2$\cite{jrc96}, 
where $\Delta_{1} < \Delta_{2} < \Delta_{3} $ are the eigenvalues of the 
symmerized shielding tensor.  It is seen that the  NMR shielding constants
separate into eight groups for carbon atom and two groups for nitrogen atoms.  
Hence, we predict that there are 10 groups of spectral signals in the NMR 
spectroscopy of ${\rm C}_{48}{\rm N}_{12}$. In contrast, the calculated 
absolute isotropic  and anisotropic NMR shielding constants for ${\rm C}_{60}$ 
are 54.10 ppm and 180.09 ppm, respectively, and there is  only one NMR spectral 
signal. In order to compare 
the predicted values of ${\rm C}_{60}$ to experimental results, we also 
list the calculated NMR shielding constants of carbon atoms in 
tetramethylsilane (TMS). We find that  the isotropic NMR shielding constant 
 of carbon atoms in ${\rm C}_{60}$, relative to that in TMS, is 146.7 ppm, 
which are in good agreement with the experimental result (=142.7\  ppm) 
of ${\rm C}_{60}$\cite{taylor90}. 
 
\begin{quote}
{\bf Table V:}  Restricted Hartree-Fock  calculations of the isotropic
$\Delta_{iso}$ and anisotropic $\Delta_{aniso}$  absolute shielding constants of the
carbon and nitrogen atoms  with a split-valence 6-31G basis set in the GIAO method 
for ${\rm C}_{48}{\rm N}_{12}$ aza-fullerene, ${\rm C}_{60}$ and 
tetramethylsilane (TMS).
\end{quote}
\begin{center}
\begin{tabular}{lcrr}\hline\hline
Center Number    &Atom  & $\Delta_{iso}$ & $\Delta_{aniso}$  \\
                 &      & [ ppm ] & [ ppm ]   \\ \hline
\{1, 13, 16, 31, 38, 51\}  & C    & 44.0  & 173.0    \\
\{4, 15, 27, 34, 40, 54\}  & C    & 57.9  & 154.9   \\
\{7, 19, 23, 43, 47, 57\}  & C    & 59.4  & 152.2   \\
\{10, 17, 25, 41, 49, 59\} & C    & 59.5  & 133.5   \\
\{2, 12, 29, 32, 37, 52\}  & C    & 67.9  & 169.2   \\
\{8, 20, 22, 44, 46, 56\}  & C    & 70.7  & 150.8   \\
\{3, 11, 28, 33, 36, 53\}  & C    & 76.6  & 141.2   \\
\{6, 18, 24, 42, 48, 58\}  & C    & 78.2  & 110.5   \\
\{9, 21, 26, 45, 50, 60\}  & N    & 95.3  & 202.6   \\
\{ 5, 14, 30, 35, 39, 55\} & N    & 114.4 & 171.5   \\
all carbons in ${\rm C}_{60}$ & C & 54.1  & 180.1  \\ 
all carbons in TMS            & C & 200.8 & 21.4     \\ \hline
\end{tabular}
\end{center}

In this work, we only focus on the split-valence 6-31G basis set and 
discuss only the frequency  analysis at 298.15 K and 1 atmosphere of 
pressure. Detailed frequency calculations and thermochemical analysis 
at different temperatures and pressures, the basis set effects including 
minimal basis sets STO-3G and polarized basis sets 6-31G*\cite{gaussian}, 
and other theoretical methods, for example, MP2\cite{gaussian} and 
CSGT method\cite{jrc96}, wll be discussed in a full paper\cite{rhxie02}.
For example,  the total electronic energies of ${\rm C}_{48}{\rm N}_{12}$ 
with respect to basis sets STO-3G, 3-21G, 6-31G and 6-31G* are 
-66795.835 eV, -67263.725 eV, -67617.316 eV and -67637.659 eV, respectively, 
which shows that improving the basis set from STO-3G to 6-31G* gives a 
significantly better total energy  and the use of polarization functions 
is not epsecially important for total energy calculations, i.e., 6-31G basis 
set can give a good estimate of the total energy of ${\rm C}_{48}{\rm N}_{12}$.

In summary, we have performed {\sl ab initio} calculation of optimized 
geometries, electronic properties, diamagnetic shielding factor, vibrational frequencies, 
IR spectrum, NMR shielding tensor, dipole polarizabilities and hyperpolarizabilities of the novel 
${\rm C}_{48}{\rm N}_{12}$ aza-fullerene. Some calculated results 
of ${\rm C}_{48}{\rm N}_{12}$ are compared with those of ${\rm C}_{60}$ at 
the level of the same theory. It is found that this novel aza-fullerene has some 
remarkable features which are different from and  can compete with 
${\rm C}_{60}$. This aza-fullerene may have  potential applications 
as semiconductor components and possible building materials for 
nanometer electronics, photonic devices and diamagnetic superconductors 
since their band gaps are small, diamagnetic shielding factor in carbon 
atom can be enhanced and electronic polarizations vary largely due to the 
doping effect.

\end{multicols}

\begin{references}
\bibitem{nist} Use of this software does not constitute an endorsement 
or certification by NIST.
\end{references}
\end{document}